\documentclass[nologo,11pt,a4paper,hidelinks]{ETHpaper}

\usepackage{graphicx}

\usepackage{booktabs}
    
\usepackage{siunitx}

\usepackage{caption}
\usepackage[authoryear]{natbib}

\usepackage{enumitem}
\setlist{nosep}
\usepackage{amsmath}

\usepackage[capitalise]{cleveref}
\usepackage[]{csquotes}
\renewcommand{\phi}{\varphi}

\title{Making Sense of Symbols: Yin and Yang in Zurich}
\titlealternative{Making Sense of Symbols: Yin and Yang in Zurich}
\author{Frank Schweitzer}
\authoralternative{Frank Schweitzer}
\address{ETH Zurich, Switzerland\\
  \url{www.sg.ethz.ch}}
\www{\url{http://www.sg.ethz.ch}}

\reference{(to be published)
}
\makeframing

\begin{document}

\maketitle

\begin{abstract}
 The widely known Yin-Yang symbol (Taijitu) is based on nested circles of different radii whose areas are colored black and white such that the interface traces an $\mathcal{S}$-shaped curve. We address the question of how this symbol can be related to physical phenomena such as daytime and nighttime duration and the annual seasons. Using a simple dynamic model of daytime duration, we introduce the excess daytime fraction and reconstruct the symbol using the latitude of Zurich. In particular, we explain how the black and white areas are linked to the stability of Yin or Yang predominance. We further demonstrate that the Golden and Silver Ratios found in the geometry of the symbol carry meaning with respect to the Gregorian calendar. Finally, we construct an alternative Yin-Yang symbol using logarithmic spirals with the Golden Ratio as the growth parameter. The didactical quantitative derivation of the Yin-Yang symbol and its grounding in real-world observations can be regarded as a novel perspective on this iconic pattern.

\end{abstract}

\section{Introduction}
\label{sec:introduction}

\begin{figure}[htbp]
  \centering
  \includegraphics[width=0.25\textwidth]{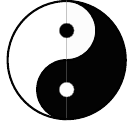}
  \caption{Recent form of the \textsc{Yin}-\textsc{Yang} symbol, denoted as \textbf{YY} henceforth.}
  \label{fig:tt}
\end{figure}

In times of trouble, conflict, and polarization, there seems to be a renewed interest in states of harmony, balance, and consensus.
The \textsc{Taijitu}, commonly known as the classic Chinese symbol for \textsc{Yin} and \textsc{Yang} (see~\cref{fig:tt}), is an iconic expression of these desired states.
It unites two opposing \enquote{forces} or \enquote{principles}, represented as \emph{black} and \emph{white}, within a circle, the ultimate symbol of wholeness and perfection.
Black and white are balanced such that each covers half of the circle's area.
Their dividing boundary, however, is not a straight line but an $\mathcal{S}$-curve, giving the symbol a dynamic impression.
Two smaller circles of opposite color resemble \enquote{eyes}, which is why some interpret the pattern as two embracing \enquote{fish}.
In earlier times the \textsc{Taijitu} had a quite different appearance\footnote{Illustrations of alternative forms can be found at \url{https://en.wikipedia.org/wiki/Taijitu}. See also \cref{fig:spirals}b.}, one that was even more reminiscent of swirling fish than today's version, which is built on nested circles (see \cref{fig:facts}a).

The widespread popularity of the \textsc{Taijitu} contrasts with the limited knowledge about its origin and meaning.
To briefly summarize its history, the \textsc{Taijitu} became popular in China only during the Song Dynasty, i.e. from the 11th century CE onward.
The philosopher \textsc{Zhou Dunyi} (1017-1073), who also wrote an insightful interpretation \citep{Wang-2005-zhoudunyis,Adler-1999-zhoudunyi},
used it to merge ideas from Daoism with Confucianism.
The concept of \textsc{Yin} and \textsc{Yang} itself originated much earlier, in the fundamental philosophy of Daoism around 500 BCE.
It plays an important role in the commentaries of the \textsc{Yijing}, the \emph{Book of Changes} \citep{Wilhelm-Baynes-1977-iching}, where the idea of a \emph{transformation} between \textsc{Yin} and \textsc{Yang} is introduced.
The complementarity of \textsc{Yin} and \textsc{Yang} and the capacity for reciprocal transformation have also inspired recent work in the quantitative sciences \citep{,Capra-1975-taophysics,Schöter-2011,Schweitzer-2022-group-i-ching}.

\textsc{Yin} and \textsc{Yang} have always been seen as two opposite yet interdependent principles.
In its earliest usage, \textsc{Yin} referred to the north and shady side of a mountain or valley, while \textsc{Yang} referred to the south and sunny side.
Later, they came to be associated with contrastive pairs such as femininity/masculinity, passivity/activity, receptive/creative, and earth/heaven.
In this paper, the connection to daytime/nighttime will be of primary relevance.
We therefore refrain from presenting a survey of the history of \textsc{Yin/Yang}.
\citet{Needham-1956-sciencecivilizationchina}, in volume 2 of his opus \emph{Science and Civilization in China}, provides background information, as does \citet{Granet-2022-chinesethought}.\footnote{Granet's book was originally published in 1934 \citep{Granet-1934-la} with a 30-page chapter on \textsc{Yin} and \textsc{Yang}, but was not translated into English until 2022.
  This is quite surprising for one of the foundational texts of Chinese structuralist anthropology and sinology.}

In the following, I refer to the \textsc{Yin}-\textsc{Yang} symbol as \textbf{YY}, for short.
To convince the reader that the famous \textbf{YY} carries meaning with respect to daytime and nighttime duration and the seasons, I begin with a simple calculation of these quantities, using \emph{Zurich} as a reference location.
The paper has a didactic aspiration: it provides all relevant equations and geometric relations, even the most elementary ones, so that readers unfamiliar with these concepts can follow the arguments.
I emphasize that the novelty of this work lies not in the relation between the \textbf{YY} and the cyclic change of seasons or day lengths, which has been mentioned in various, and partly opaque, contexts \citep{Ivankovic-2020-astronomicalinaccuracy,Leong-2024-daoquantummechanics}.
Rather, the aim is to calculate how such relations materialize in a physical, or astronomical, setting.
This quantification then forms the basis for exploring how prominent proportions such as the Golden Ratio and the Silver Ratio manifest themselves in a concrete manner.
Finally, insights from the Golden Ratio are used to construct an alternative pattern for the \textsc{Yin}-\textsc{Yang} symbol, based on Golden Spirals.

\section{Notable Features and Open Questions}
\label{sec:facts}

\begin{figure}[htbp]
  \centering
  \includegraphics[width=0.36\textwidth]{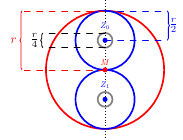}(a)
  \hspace*{1cm}
  \includegraphics[width=0.28\textwidth]{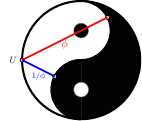}(b)
    \caption{Geometry of the \textsc{Yin-Yang} symbol (\textbf{YY}): (a)  Construction using nested circles of different radii, (b) Black/white filling of the areas resulting from (a) and identification of distances obeying the Golden Ratio, $\phi$.}
  \label{fig:facts}
\end{figure}

There is vast, and quite often repetitive, information about the \textbf{YY} scattered across the world wide web.
In the following I focus on the geometric structure only and summarize two basic features of the symbol.
The first regards the geometric construction of the \textbf{YY} \citep{Browne-2007-taiji}.
Its \emph{modern} version is composed of nested circles, as \cref{fig:facts}(a) shows.
It lacks the swirling impression of ancient versions (see \cref{fig:spirals}b), but has the advantage of a straightforward procedure to generate the pattern.

We start with a (red) unit circle of radius $r=1$ and center point $\mathrm{M}$.
Inside this circle, two smaller (blue) circles of radius $r/2$ and center points $\mathrm{Z}_{0}$, $\mathrm{Z}_{1}$ are inscribed.
They intersect at $\mathrm{M}$ and are each internally tangent to the outer circle.
Inside each of these two circles, a (grey) circle of even smaller radius is placed.
We have used $r/8$ here, so the diameter of the smallest circles is $r/4$.
Variations in their size are also common.
These two smallest circles are referred to as ``eyes'' in the following.

Once the geometric structure is established, the areas resulting from the intersections of the different circles with the vertical axis are filled \emph{black or white} such that the familiar appearance of the \textbf{YY} results.
The size of the black and white areas is obviously half the size of the unit circle because of the underlying symmetry.

A second notable feature regards the lengths of two prominent distances inside the \textbf{YY}.\footnote{According to \url{https://www.cut-the-knot.org/do_you_know/GoldenRatioInYinYang.shtml}, this finding should be attributed to John Arioni. The geometric proof is straightforward. Yet, the meaning of the Golden Ratio with respect to the \textbf{YY} remains unexplained.}
We start from a point $\mathrm{U}$ on the perimeter of the unit circle that is equidistant from the center points of the two smaller circles.
\cref{fig:facts}(b) shows one of the two symmetric possibilities for locating $\mathrm{U}$.
Then, drawing a line from $\mathrm{U}$ through each of the center points $\mathrm{Z}_{0}$, $\mathrm{Z}_{1}$, these lines cross the border between the black and the white area at different distances.
Because the outer circle is the unit circle,
one can verify that these two distances are related by the \emph{Golden Ratio}, i.e. one of them (red) has the length $\phi=1.618$ and the other one (blue) the length $1/\phi=0.618$.

While these conclusions are straightforward, their \emph{interpretation} remains unclear.
We address three questions in this paper:
\begin{enumerate}
\item What is the \emph{meaning} of the outer circle, the two inner circles, and the two eyes? Are they purely geometric shapes, or do they reflect something real?
\item What is the \emph{meaning} of the black and white fillings? Does the prominent ``fish-like'' shape have a foundation in reality?
\item How should the \emph{Golden Ratio} within the \textbf{YY}  be interpreted? Is it a purely geometric artifact, or is it related to physical quantities?
\end{enumerate}
In essence, these three questions ask whether the \textbf{YY}  should be understood as an abstract geometric representation of philosophical concepts or as a reflection of real observations.
These two perspectives are not mutually exclusive; both have their own validity.

In the following, I will demonstrate that the \textbf{YY}  can indeed be linked to real observations.
Interpreting these observations aligns nicely with the philosophical perspective.
Hence, these investigations not only provide a didactic exercise in generating the \textbf{YY}  from data,
but also shed new light on the scope of classical Chinese concepts.

\section{Daytime and nighttime in Zurich}
\label{sec:daylight}

\subsection{Calculating the daylight duration}
\label{sec:calc-dayl-durat}

This section outlines in a didactical manner how to derive the \textbf{YY} from data on daylight duration.
There are tables available online.\footnote{\url{https://www.timeanddate.de/sonne/schweiz/zuerich?month=6&year=2026}}
We use an analytic approach here because the rotation of the Earth over 24 hours and its motion around the sun are quite regular.
Only two parameters play a role: $\beta=23.44^{\mathrm{o}}$ is the axial tilt, or obliquity, of the \emph{Earth}, measured in degrees.
$\lambda = 47.37^{\mathrm{o}}$ is the geographic latitude, in degrees, of our \emph{location},
for which we take the value of Zurich, Switzerland.
Hence, all subsequent calculations about daytime and nighttime reflect the situation in \emph{Zurich}.
However, they can easily be repeated for any other location by simply substituting the respective latitude $\lambda$.

The daylight duration in hours, $D$, is the time between sunrise and sunset.
Considering the altitude of the sun, or solar elevation, i.e. the angle of the sun above the horizon, $h$,
at sunset and sunrise, $h=0$ by definition.
As the earth rotates $15^{\mathrm{o}}$ per hour, $h$ increases until the \emph{local solar noon}, where the sun is on the meridian, i.e. the longitude $8.54^{\mathrm{o}}$ east for Zurich.
Afterwards it decreases again.
Following textbooks on practical astronomy\footnote{The equations have not changed over the last hundred years, so it is fair to cite a classic \citep{Nassau-1932-textbookpracticalastronomy}. Modern handbooks, e.g. \citep{Meeus-1991-astronomicalalgorithms}, provide complex algorithms that are not needed here.}
its value measured on a daily timescale $\tau$ is expressed as
\begin{align}
  \label{eq:10}
\sin h[\tau,\delta(t)]= \sin\lambda \,\sin\delta(t) + \cos\lambda \,\cos\delta(t) \,\cos H[\tau,\delta(t)]
\end{align}
where $H$ is the \emph{hour angle}, measured in degrees $(^{\mathrm{o}})$.
It indicates how far in time sunrise or sunset is from local solar noon, expressed as an angle rather than in hours.

$H$ depends, in addition to the time of day, also on the solar declination $\delta(t)$, i.e. the angle between the sun and the earth's equator.
$t$ is the day of the year, numbered from 1 to 365.
$\delta(t=T)=0$ during the vernal and autumnal equinoxes, i.e. at particular day numbers $t=T$ during the year, and reaches its maximum value at the summer and winter solstices (see also \cref{tab:calendar}).
\begin{align}
  \label{eq:13}
   \delta(t) &= \beta \cdot \sin\left[\frac{360^{\mathrm{o}}}{365}(t - T)\right]  
\end{align}
The fraction $360^{\mathrm{o}}/365$ is used to map time $t$ in days to degrees on the unit circle with $360^{\mathrm{o}}$.

To obtain the value of the hour angle $H$ at sunrise/sunset, we use $h=0$ and solve \cref{eq:10} for $H=H_{0}$. 
\begin{align}
  \label{eq:11}
    H_0[\delta(t)] &= \arccos\left[-\tan\lambda \cdot \tan\delta(t)\right]
\end{align}
$H_{0}$ is the sunrise/sunset hour angle.
By convention, $H = 0^\circ$ at the local solar noon.
At sunrise $H = -H_0$ is negative, at sunset $H = +H_0$ is positive.
This defines our daylight duration.
Going from $-H_{0}$ to $+H_{0}$ at $15^{\mathrm{o}}$/hour, the daylight duration is
\begin{align}
  \label{eq:12}
  D[\delta(t)] &= \frac{2}{15^{\mathrm{o}}/\mathrm{hour}}\, H_{0}[\delta(t)]=\frac{24 \mathrm{\,h}}{180^{\mathrm{o}}}\, H_0[\delta(t)]
\end{align}
The day number $T=81$ in \cref{eq:13} is used to normalize $D$ such that it is precisely $D_{\mathrm{eq}}=12$ hours, i.e. half of the day, at the vernal equinox, as shown in \cref{tab:calendar}.

To visualize the solar year of 365 days on a unit circle, we need to map the day number $t$ to a degree $\gamma$ on the circle.
Defining $\gamma=0$ for the winter solstice, the new year starts 11 days after the winter solstice, i.e. $\gamma$ has to be calculated as:
\begin{align}
  \label{eq:14}
  \gamma&=\frac{360^{\mathrm{o}}}{365}(t+10)\;;\quad
  t=\frac{365}{360^{\mathrm{o}}}\gamma-10
\end{align}
Then the solar declination w.r.t $\gamma$ reads
\begin{align}
  \label{eq:13a}
   \delta(\gamma) &= \beta \cdot \sin\left[\gamma-90^{\mathrm{o}}\right] = -\beta \cos \gamma  
\end{align}

\begin{table}[htbp]
  \centering
  \begin{tabular}[c]{S[table-format=2.2] rr S[table-format=2.2] S[table-format=2.2] cl}
    \toprule
    {Date} & {$t$} & {$\gamma$ ($^{\mathrm{o}}$)} & {$\quad\delta$ ($^{\mathrm{o}}$)} &{$D$ (hours)} & {Symbol} & {Meaning} \\
    \midrule
1.1 & 1 & 11   & -23.00 & 8.34 & & new year \\    
    21.3 & 81 & 90 & 0 & 12.00 & V &vernal equinox \\
    21.6 & 172 & 180 & 23.44 & 15.75 & S& summer solstice \\
    21.9 & 264 & 270 & 0 & 12.00& U& autumn equinox \\
    21.12 & 355 & 0 & -23.44 & 8.25 & W& winter solstice \\
    \midrule
    24.4 & 114 & 122 & 12.61 &  13.87& &\\
    24.11& 328 & 333  & -21.00 & 8.71& & \\
    \bottomrule
  \end{tabular}
  \caption{Date (Gregorian), day number $t$, circular position $\gamma$, \cref{eq:14}, solar declination $\delta$, \cref{eq:13a}, daylight duration $D$, \cref{eq:12}, for Zurich . Symbols are used for cardinal points during the solar year in subsequent figures.}
  \label{tab:calendar}
\end{table}

\cref{tab:calendar} gives an overview of the resulting values in a stylized manner.
Because the simplified equations above assume a very regular motion of the Earth, the dates of the astronomical seasons (the two solstices and the two equinoxes) are spread evenly throughout the year.
A quarter year is rounded to 91 full days.
Since the winter solstice is fixed at 21.12., which is day 355, the other cardinal points are given by day numbers 81, 172, and 264.
The resulting small errors and deviations from true astronomical measurements are discussed in a subsequent section at the end.

\begin{figure}[htbp]
  \centering
  \includegraphics[width=0.43\textwidth]{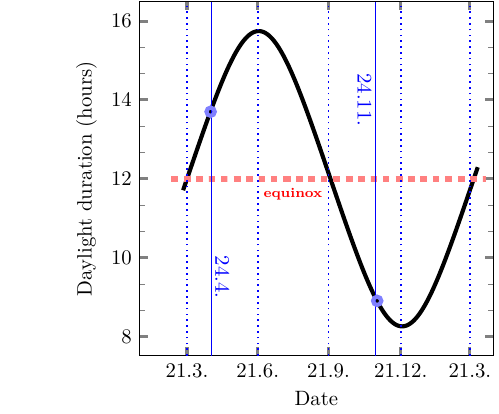}(a)\hfill
    \includegraphics[width=0.5\textwidth]{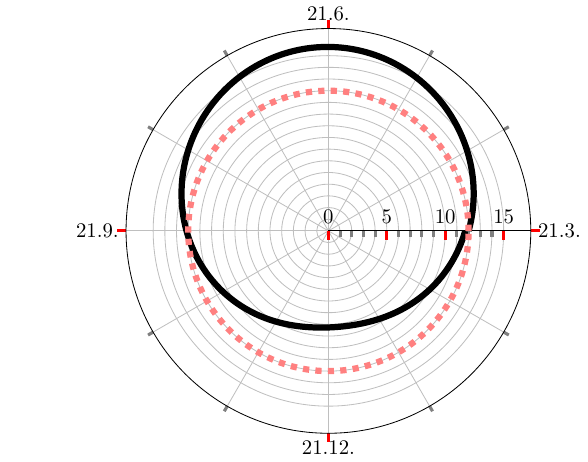}(b)
  \caption{Daylight duration for Zurich, $\lambda=47.37^{\mathrm{o}}$. (a) $D$, \cref{eq:12} on a linear time scale and (b) using polar coordinates. The dashed red line indicates $D=12$ h.}
  \label{fig:linear}
\end{figure}

\cref{fig:linear}(a) visualizes the change in daylight duration for Zurich over the solar year, as obtained from \cref{eq:12}.
In addition to the two equinoxes ($t=81$ and $t=264$) with $D_{\mathrm{eq}}=12$ h, we verify that for Zurich $D_{\mathrm{max}}=15.75$ h at the summer solstice ($t=172$) and $D_{\mathrm{min}}=8.25$ h at the winter solstice ($t=355$).
Note that $D_{\mathrm{max}}+D_{\mathrm{min}}=24$ h.

Because the oscillation of the daylight duration is clearly visible, it makes sense to introduce polar coordinates.
Time $t$ is mapped to the angle $\gamma$, \cref{eq:14}.
In \cref{fig:linear}(b), we have placed the winter solstice at the bottom ($\gamma=0$) and the summer solstice at the top ($\gamma=180$), with time running counter-clockwise.
Daylight duration $D$ is mapped to the radius of the polar coordinates.
The asymmetric upward shift of the black curve in \cref{fig:linear}(b) reflects the latitude of Zurich, which is not located on the equator.
Over the course of the year, days are longer than 12 hours between spring and fall, and shorter than 12 hours between fall and spring.

In a final step, we normalize the daylight duration to the interval $[0,1]$ introducing a reduced radius $\rho$ as follow:
\begin{align}
  \label{eq:15}
  \rho(\gamma) =
  \begin{cases}
    \dfrac{D(\gamma) - D_{\mathrm{eq}}}{D_{\max} - D_{\mathrm{eq}}} & \text{if } D(\gamma) \geq D_{\mathrm{eq}} \\[18pt]
    \dfrac{D_{\mathrm{eq}}-D(\gamma)}{D_{\mathrm{eq}}- D_{\min}} & \text{if } D(\gamma) < D_{\mathrm{eq}}
  \end{cases}
\end{align}

\begin{figure}[htbp]
  \centering
  \includegraphics[width=0.45\textwidth]{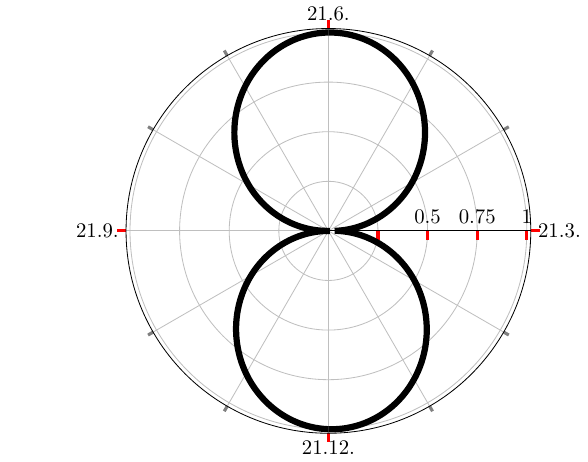}(a)\hfill
    \includegraphics[width=0.45\textwidth]{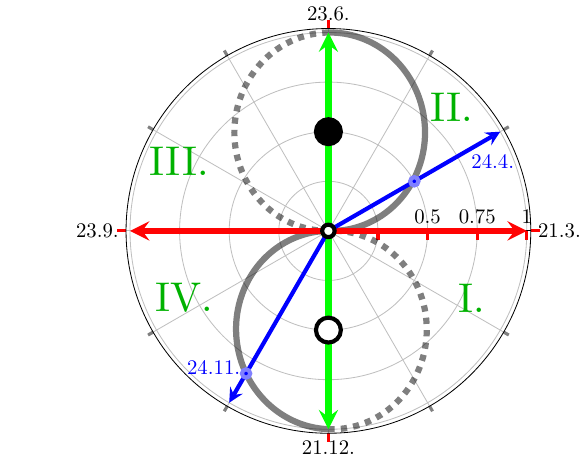}(b)
  \caption{(a) Replot of \cref{fig:linear}(b) using the reduced variable $\rho(\gamma)$, \cref{eq:15}, instead of the daylight duration $D(\gamma)$, \cref{eq:12}. (b) Auxiliary lines for two arbitrary dates, 21.4, 21.11.}
  \label{fig:polar}
\end{figure}

This definition of $\rho(\gamma)$ uses the fact that oscillations are symmetric around the value $D_{\mathrm{eq}}=12$ h.
Instead of positive and negative deviations from $D_{\mathrm{eq}}$, only their absolute values are taken into account.
The result is shown in \cref{fig:polar}(a), which should be compared with the geometric construction in \cref{fig:facts}(a).
We see that within the outer circle two smaller circles have emerged, which are already reminiscent of the geometric appearance of the YY symbol.
But now, these two smaller circles have a \emph{meaning}, as they visualize the values of $\rho(\gamma)$.
This will be explored in more detail in the following.

\subsection{Explaining the excess curves}
\label{sec:expl-excess-curv}

By definition, $\rho(\gamma)$ measures \emph{deviations} from the maximum or minimum daylight duration, normalized to the maximum possible difference, which for Zurich is:
\begin{equation}
  \label{eq:9}
  \begin{array}[c]{rclclcl}
    \Delta D & = & D_{\mathrm{max}}-D_{\mathrm{eq}}&= & 15.75\mathrm{h}-12\mathrm{h} &= & 3.75 \mathrm{h} \nonumber \\
             & = & D_{\mathrm{eq}}-D_{\mathrm{min}}& =& 12\mathrm{h} - 8.25\mathrm{h} &= & 3.75 \mathrm{h}
  \end{array}
\end{equation}
Hence, $\rho(\gamma)$ denotes the \emph{excess fraction} of the daytime or nighttime, respectively, when $D(\gamma)<D_{\mathrm{eq}}$.
To better understand the meaning of the curve $\rho(\gamma)$, we use the auxiliary \cref{fig:polar}(b).
In this polar diagram, \emph{time}, i.e. degrees, runs counter-clockwise on the outer circle.

The solar year is divided into four \emph{quarter years}, labeled I. to IV.
In this diagram, pointers of different colors are shown.
The \emph{angle} of a pointer from the center to the perimeter indicates a \emph{date}, while the intersection of the pointer with one of the inner circles gives the value of $\rho$.

In the polar diagram the (red) pointers towards the two equinoxes are on either side of the \emph{horizontal} axis, whereas the (green) pointers towards the summer and winter solstices are on either side of the \emph{vertical} axis.
We verify that the red pointers intersect with the $\rho(\gamma)$ curve only at zero, i.e. $\rho=0$ at the equinoxes, by definition.
The green pointers intersect with the $\rho(\gamma)$ curve at one.
By definition, the daylight duration reaches its maximum $D_{\mathrm{max}}$ at the summer solstice and its minimum $D_{\mathrm{min}}$ at the winter solstice, both of which are mapped to $\rho=1$.

Now we choose a spring date in the II. Quarter, e.g. 24.4. (see also \cref{tab:calendar}).
The (blue) pointer intersects with the inner circle at $\rho=0.5$.
This denotes the \emph{excess fraction} of the daytime with respect to the difference $\Delta D=3.75$h, and 50\% of it is $1.87$ h, or 1 hour and 
52 minutes. 
So, the actual daylight duration on that date is 13.87 hours.

The example illustrates that the curve $\rho(\gamma)$, which generated the two inner circles, has a clear interpretation as the \emph{excess fraction}.
This also works for dates in the fall, e.g. 24.11. in the IV. Quarter. 
The (blue) pointer in \cref{fig:polar}(b) intersects with the inner circle, now in the lower half of the outer circle, i.e. there is an excess in \emph{nighttime}. 
We find $\rho=0.87$, so it is 87\% of $\Delta D=3.75$h,
i.e. 3.26h or 3 hours and 16 minutes. 
These must now be \emph{subtracted} from $D_{\mathrm{eq}}$, giving a daylight duration of 8 hours and 44 minutes.

To conclude, the two inner circles in the \textbf{YY} have a clear \emph{physical} interpretation as fractions of the daytime/nighttime differences between equinoxes and solstices.\footnote{It should be noted that my conclusion, namely that the \textbf{YY} \emph{accurately} reflects physical observations, contradicts \citet{Ivankovic-2020-astronomicalinaccuracy}, who claims its \emph{inaccuracy}, albeit using a different approach not providing detailed calculations.}

In principle, this evaluation also works for dates between the summer solstice and the autumn equinox, III. Quarter, or between the winter solstice and the vernal equinox, I. Quarter.
The intersections will then lie on the \emph{dashed} arcs of the inner circles, shown in \cref{fig:polar}(b).
But, as we can verify by a comparison with the \textbf{YY}, \cref{fig:tt}, these intersections \emph{do not} play a role in constructing the black/white interface of the $\mathcal{S}$-curve.
Only the \emph{solid} arcs of the inner circles are taken into account.
This raises a very interesting question: Why are the dashed parts of the inner circles, despite having a physical meaning, \emph{not} used in the \textbf{YY}?

\subsection{Stability}
\label{sec:stability}

The answer to our question: the \textbf{YY}  does not simply reflect the actual values of the daytime/nighttime duration, in terms of the excess fractions.  
Instead it indicates under what conditions the subtle prevalence of \textsc{Yin} or \textsc{Yang} changes.
In other words, this prevalence is \emph{not} identical to the daytime/nighttime relation.
Instead it reflects the \emph{stability} of the balance.

\begin{figure}[htbp]
  \centering
  \includegraphics[width=0.28\textwidth]{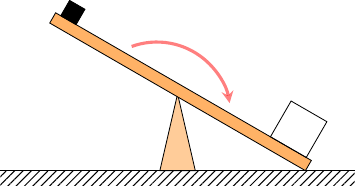}(a)\hfill
  \includegraphics[width=0.28\textwidth]{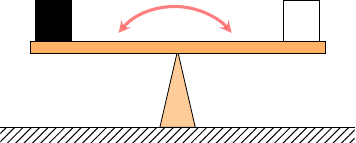}(b)\hfill
  \includegraphics[width=0.28\textwidth]{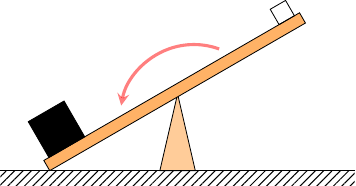}(c)
  \caption{Seesaw analogy: The seesaw will only move if the weight on the left becomes larger than the weight on the right, and vice versa. (a) $x(t)>y(t)$,
    (b) $x(t)=y(t)$, Vernal and autumn equinoxes. (c) $x(t)< y(t)$.   }
  \label{fig:wippe}
\end{figure}
In order to see this, I use an analogy everyone knows from childhood: a seesaw (see \cref{fig:wippe}).
It has weights on either side of the board and is supported by a pivot point in the middle.
The two weights shall represent \textsc{Yin} (black), as $y(t)$, and \textsc{Yang} (white) as $x(t)$.
Their box sizes indicate their ``weight'' relative to the other side, i.e. $x(t)+y(t)=1$, the sum being constant.
The seesaw is clearly in balance only if $x(t)=y(t)$, which is the mechanical analogy of the autumn and vernal equinoxes (see \cref{fig:wippe}b).

This equilibrium, however, is \emph{unstable}, meaning that a small excess of weight on one side causes the seesaw to move.
The seesaw differs from a weighing scale in that its pivot point is located \emph{below} the center of mass.
A scale, by contrast, has the center of mass \emph{above} the pivot point, which allows it to
settle into a new equilibrium after the weights have changed.
In contrast, any small perturbation of the equilibrium of a seesaw will cause it to tip, and
the motion will only stop once one side of the board hits the ground.
That is the only stable equilibrium.

\begin{figure}[htbp]
  \centering
  \includegraphics[width=0.4\textwidth]{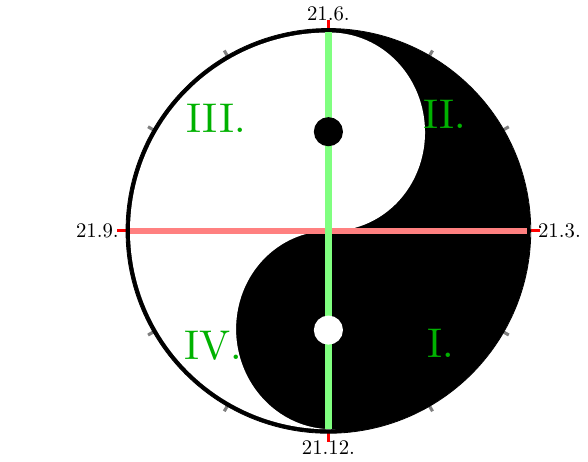}(a)\hfill
    \includegraphics[width=0.45\textwidth]{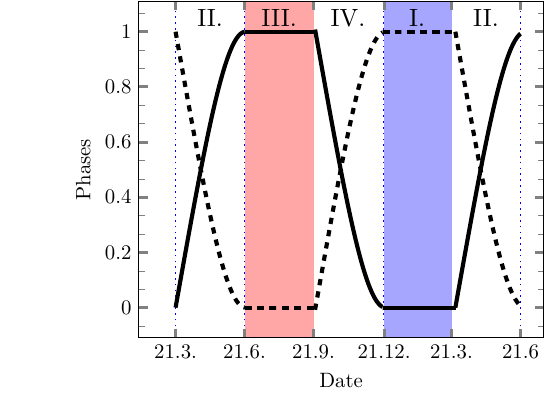}(b)
  \caption{(a) \textsc{Yin}-\textsc{Yang} symbol (\textbf{YY}) with quadrants of stability (see text). (b) Predominance of \textsc{Yin} (dashed line) or \textsc{Yang} (solid line). (white) Transition phases, (blue/red) Metastable \textsc{Yin}/\textsc{Yang} phases.}
  \label{fig:filled}
\end{figure}
Taking our example of the daytime/nighttime duration, the relative weights $x(t)$ and $y(t)$ change because of an external dynamic, i.e. the earth's revolution around the sun.
Therefore, the unstable equilibrium at the vernal equinox is left behind.
As \cref{fig:filled} shows,
during the II. Quarter, daytime, i.e. $x(t)$, increases and \textsc{Yang} gains, which causes the seesaw to turn.
It only stops at the summer solstice, where $x(t)$ reaches its maximum.
In the polar diagram, this is reflected in the excess fraction $\rho(\gamma)$, increasing from zero to one during the II. Quarter.

$\rho=1$ would be the final state \emph{if} nothing caused \textsc{Yang} to decrease and \textsc{Yin} to increase.
But precisely at this point, the pointer towards the summer solstice at the top
crosses the small black ``eye'' of \textsc{Yin}.
This can be seen as a seed for \textsc{Yin} to grow again.
If $y(t)$ grows, $x(t)$ must shrink, which happens in the III. Quarter of the year.

But will the seesaw tilt?
The answer is: Not yet, because the seesaw can only tip once $y(t)>x(t)$, that is \emph{after} the autumn equinox, during the IV. Quarter. 
That means, during the III. Quarter \textsc{Yang} already shrinks and \textsc{Yin} grows, but nothing happens because the seesaw is still in its stable state.
Only after the equilibrium $y(t)=x(t)$ is passed does the seesaw start tilting again, and it continues to do so until the next stable state is reached at the winter solstice, where \textsc{Yin} has grown to the maximum and $\rho=1$.

In this situation, however, the pointer crosses the small white ``eye'' of \textsc{Yang}, which is a seed for \textsc{Yang} to grow anew. 
During the I. Quarter, \textsc{Yin} shrinks and \textsc{Yang} grows, but the seesaw does not move yet, because it is still in its stable state.
Only after the vernal equinox is reached does the seesaw start tilting, and it continues to do so during the II. Quarter until the next stable state, the summer solstice, is reached.

This allows us to give a precise explanation of why the different Quarters of the year in \cref{fig:filled}(a) are colored black and white the way they are.
The \emph{color} indicates the \emph{stability} of the given state, whereas the black/white interface of the $\mathcal{S}$-curve indicates the \emph{change} of that state.
The I. and III. Quarters are entirely black or white, because the prevalence of \textsc{Yin} or \textsc{Yang} is stable, albeit decreasing.
This is the \emph{stable} situation of the seesaw, not tilting.
The II. and IV. Quarters show the $\mathcal{S}$-curve, because there is no stable dominance of either \textsc{Yin} or \textsc{Yang}.
This is the \emph{unstable} situation of the seesaw, tilting.

\cref{fig:filled}(b) shows the same dynamics using a linear time scale instead of polar coordinates.
We can clearly see that phases of \enquote{tranquility} (1st and 3rd quarters) alternate with phases of \enquote{activity} (2nd and 4th quarters).
This is in perfect agreement with an 11th century description by \textsc{Zhou Dunyi}, who wrote:
\enquote{Activity and stillness alternate; each is the basis of the other} \citep{Adler-1999-zhoudunyi}.

\section{Analytic expressions}
\label{sec:analytic-expressions}

\subsection{Metallic ratios}
\label{sec:metallic-ratios}

Based on our interpretation of the $\mathcal{S}$-Curve, we can now address the third research question about the meaning of the Golden Ratio in the \textbf{YY}.

The Golden Ratio is the most prominent representative of a whole class of proportion measures, including the Silver Ratio, the Bronze Ratio, and other ``metallic'' ratios \citep{Spinadel-1999-metallicmeans,Dunlap-1997-goldenratio,Schroeder-1991-numbertheory}.\footnote{See also \url{https://en.wikipedia.org/wiki/Metallic_mean}}

The geometric interpretation assumes a unit of length $L$ that is divided into a longer segment, $a$, and a shorter segment, $b$.
The ratio $a/b$, $a>b$, must satisfy a different condition for each ``metallic'' mean:

\begin{align}
  \label{eq:1}
  \frac{a}{b}&=\frac{L}{a}=\frac{a+b}{a}=\phi \\
  \label{eq:2}
    \frac{a}{b}&=\frac{L}{a}=\frac{2a+b}{a}=\sigma 
\end{align}

$\phi$ is the Golden Ratio if the relation between the \emph{longer} segment $a$ and the shorter segment $b$ is the same as the relation between the length $L$ and the longer segment $a$.
In essence it is a condition for \emph{self-similarity}.
For the Silver Ratio $L$ is divided differently, namely into \emph{two} portions of $a$ and \emph{one} portion of $b$.
For a Bronze Ratio, rarely used, $L$ would be divided into $3a$ and $b$, and so forth.

The original measures play a role, e.g., in architecture where segments or areas need to be divided according to certain proportionality rules.
For the \textbf{YY}, their meaning remains unclear.
Of course, we can define geometrical distances and analyze their relations.
But if we detect a Golden Ratio, we still lack its interpretation with respect to the \textbf{YY}.

Therefore, it is worth choosing a different starting point.
Consider two quantities $x$ and $y$ that are related to one another in an ambiguous manner:
\begin{subequations}
  \begin{align}
    \label{eq:3}
    y&=-\frac{1}{x}\;; \quad x+y=n\\
    \label{eq:4}
    x&=n +\frac{1}{x}\\
    \label{eq:5}
    0&=x^{2}-n x-1
  \end{align}
\end{subequations}
The condition \cref{eq:3} requires that $x$ and $y$ have opposite signs.
Their absolute values are reciprocal, i.e., the larger $x$ is, the smaller $y$ is.
At the same time the sum of $x$ and $y$ is fixed, where $n$ is a natural number.
So $x$ and $y$ represent two complementary quantities subject to opposing forces.
Together they form a unit $n$, but the increase of one comes at the expense of the other.
\textsc{Yin} and \textsc{Yang} would be fitting candidates for such a relationship.

The quadratic \cref{eq:5} can be solved:
\begin{subequations}
  \begin{align}
    \label{eq:6}
    x_{1,2}&=\frac{n}{2} \pm \frac{n}{2} \sqrt{n^{2}+4} \\
    x_{1,2}&=\frac{1}{2} +\frac{1}{2} \sqrt{5} \quad \mathrm{if}\ n=1\\
    x_{1,2}&=\frac{2}{2} +\frac{2}{2} \sqrt{8} = 1+ \sqrt{2} \quad \mathrm{if}\ n=2 \\
    x_{1,2}&=\frac{3}{2} +\frac{3}{2} \sqrt{13} = 1+ \sqrt{2} \quad \mathrm{if}\ n=3
  \end{align}
\end{subequations}
If the natural number $n$ is equal to one, the solutions for $x$ give us the \emph{Golden} Ratio, for $n=2$ the \emph{Silver} Ratio, and for $n=3$ the \emph{Bronze} Ratio.  
It is worth noting that for $x$ we take only the positive solutions into account.
Solutions for $n\geq 3$ are not discussed in the following.

The algebraic solutions alone already tell us that the Golden Ratio is related to the \emph{pentagon}/\emph{pentagram} because of $\sqrt{5}$, while the Silver Ratio is related to the \emph{octagon} because of $\sqrt{8}$ and to the \emph{square} because of $\sqrt{2}$.

\subsection{Golden Ratio}
\label{sec:golden-ratio}

\begin{figure}[htbp]
  \centering
  \includegraphics[width=0.3\textwidth]{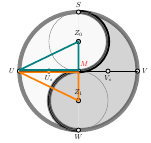}(a)\hfill
  \includegraphics[width=0.3\textwidth]{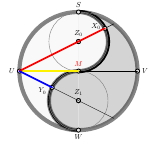}(b)\hfill
  \includegraphics[width=0.3\textwidth]{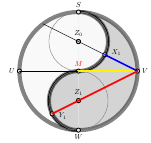}(c)
  \caption{\textsc{Yin}-\textsc{Yang} symbol (\textbf{YY})  with cardinal points (see \cref{tab:calendar}): (a) Right triangles to calculate distances, (b) Golden ratios $\phi$ (red), $1/\phi$ (blue), \cref{eq:gs-40}, and reference segment $\overline{\mathrm{UM}}$ (yellow), (c) Same as (b), but for reference segment $\overline{\mathrm{VM}}$.}
  \label{fig:golden}
\end{figure}

The solution for the Golden Ratio  shows some interesting properties:
\begin{subequations}
  \begin{align}
    \label{eq:7}
    x_{1}&= \frac{1}{2} + \frac{1}{2} \sqrt{5} = \phi= 1.618 \\
  \phi -1 &= \frac{1}{\phi} = \frac{2}{1+\sqrt{5}}=\frac{\sqrt{5}-1}{2} =0.618
\end{align}
\end{subequations}
In order to explain the Golden Ratio  in the \textbf{YY}, we add two auxiliary triangles  shown in \cref{fig:golden}(a), to identify intersections and distances. 

The anchor points for the geometric analysis are the equinoxes $\mathrm{V}$ and $\mathrm{U}$ and their connections through the centers of the ``eyes'', $\mathrm{Z}_{0}$, $\mathrm{Z}_{1}$.
Starting from $\mathrm{U}$ and the large circle with radius $r=1$, we can identify in \cref{fig:golden}(a) \emph{two triangles}, $(\mathrm{U,M,Z_{0}})$ and $(\mathrm{U,M,Z_{1}})$.
Because of $\overline{\mathrm{UM}}=1$ and $\overline{\mathrm{Z_{0}M}}=\overline{\mathrm{Z_{1}M}}=1/2$ by construction, \cref{fig:facts}(b),
we verify from the Pythagorean theorem, $a^{2}+b^{2}=c^{2}$, that $\overline{\mathrm{UZ_{0}}}=\overline{\mathrm{UZ_{1}}}=\sqrt{5}/2$.
This gives us the distances of interest, $\overline{\mathrm{UX_{0}}}=1.618$ and $\overline{\mathrm{UY_{0}}}=0.618$, as follows:
\begin{equation}
    \label{eq:gs-40}
    \begin{aligned}
      \overline{\mathrm{UX_{0}}}&=\overline{\mathrm{UZ_{0}}}+\overline{\mathrm{Z_{0}X_{0}}}
= \frac{\sqrt{5}}{2}+\frac{1}{2}=\phi  \\
      \overline{\mathrm{UY_{0}}}&=\overline{\mathrm{UZ_{1}}}-\overline{\mathrm{Z_{1}Y_{0}}}
= \frac{\sqrt{5}}{2}-\frac{1}{2}=\phi-1=\frac{1}{\phi} 
    \end{aligned}
  \end{equation}
Reproducing the known result about the \emph{length} of these two distances was the easy part.
  The more challenging question is about their meaning.

  We recall that the Golden Ratio describes a relation between \emph{two segments}.
  In our case, the reference segment is $\overline{\mathrm{UM}}=1$ (in yellow).
  A Golden Ratio is obtained in \emph{two} cases (see \cref{fig:golden}b):
  \begin{enumerate}[itemsep=0em]
  \item $\overline{\mathrm{UM}}=1$ is the \emph{shorter} segment and $\overline{\mathrm{UX_{0}}}=\phi=1.618$ (in red) is the \emph{longer}
    segment.
    We call this the \emph{outer solution} in the following.
  \item $\overline{\mathrm{UM}}=1$ is the \emph{longer} segment and $\overline{\mathrm{UY_{0}}}=1/\phi=0.618$ (in blue) is the \emph{shorter}
    segment.
    We call this the \emph{inner solution} in the following.
  \end{enumerate}
Furthermore, we can repeat the calculation for the other cardinal point, $\mathrm{V}$, to mirror the results obtained for $\mathrm{U}$ (see \cref{fig:golden}c).
Hence, in the case of the Golden Ratio, we have \emph{four} points of interest for our further analysis, all satisfying the Golden Ratio condition: the \emph{outer} solutions
$\mathrm{X}_{0}$, $\mathrm{Y}_{1}$, and the inner solutions $\mathrm{X}_{1}$, $\mathrm{Y}_{0}$.
In the following, we use the cardinal point $\mathrm{U}$.
The solutions for $\mathrm{V}$ do not change the results, but the angles must be renormalized.

\begin{figure}[htbp]
  \centering
  \includegraphics[width=0.5\textwidth]{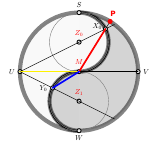}
  \caption{Interpretation of the Golden Ratio in the \textsc{Yin}-\textsc{Yang} symbol (\textbf{YY}): Definitions of $\mathrm{X}_{0}$, $\mathrm{Y}_{0}$ and $\mathrm{P}$.}
  \label{fig:gold}
\end{figure}
In order to find an interpretation for the points $\mathrm{X}_{0}$ and $\mathrm{Y}_{0}$ on the $\mathcal{S}$-curve, we remind their meaning with respect to the center $\mathrm{M}$ which defines the ratio $\rho$.
So we have (see \cref{fig:gold})
\begin{align}
  \label{eq:17}
  \rho_{X_{0}}&=\overline{\mathrm{MX_{0}}} = 0.850 \nonumber \\
  \rho_{Y_{0}}&=\overline{\mathrm{MY_{0}}} = 0.525 \nonumber \\
  \frac{\rho_{X_{0}}}{\rho_{Y_{0}}}&=1.619 
\end{align}
It should be noted that because of symmetry $\overline{\mathrm{MY_{0}}}=\overline{\mathrm{SX_{0}}}$,
so the two radii $\rho_{X_{0}}$ and $\rho_{Y_{0}}$ form a right triangle in the inner circle, and\begin{align}
  \label{eq:18}
  \rho_{X_{0}}^{2}+\rho_{Y_{0}}^{2}=1
\end{align}
To conclude, the Golden Ratio shown in \cref{fig:facts}(b) defines \emph{two points} on the $\mathcal{S}$-Curve, $\mathrm{X}_{0}$ and $\mathrm{Y}_{0}$.
The distances of these two points fulfill the conditions of the Golden Ratio
with respect to $\mathrm{U}$ (the inner and outer solutions), but also with respect to $\mathrm{M}$, because they define two excess fractions, $\rho_{X_{0}}$ and $\rho_{Y_{0}}$, that are related by the Golden Ratio.
Hence, the Golden Ratio
has a physical interpretation in terms of daytime and nighttime duration.

In addition to the Golden Ratio of the excess fractions, there is another remarkable observation.
Starting from the center point $\mathrm{M}$, the Golden Ratio point $\mathrm{X}_{0}$ can be projected onto the outer circle, which contains the days of the year (see \cref{fig:gold}).
The projection yields a point $\mathrm{P}$ that, as we can verify, has the angle $\gamma=148.5^{\mathrm{o}}$.
Using \cref{eq:14}, we can transform this angle into a day of the year, which gives day 140.
Interestingly, this date divides the solar year of 365 days according to the Golden Ratio:
\begin{align}
  \label{eq:16}
  \frac{a}{b}=\frac{365}{225}=1.622 \;;\quad  \frac{a+b}{a}=\frac{590}{365}=1.616 \;;\quad 
  \frac{a}{b}=\frac{225}{140}=1.607   %
  \end{align}
All values are very close to $\phi=1.618$.

So, we can conclude that the Golden Ratio has a real interpretation with respect to the \textbf{YY}.
First, it defines the ratio of the excess fractions that have a physical meaning for the daylight duration.
Secondly, it defines a date in the solar year such that the number of days elapsed and the number of days remaining obey the Golden Ratio.

\subsection{Silver Ratio}
\label{sec:silver-ratio}

It is interesting to see whether the above procedure also works for the Silver Ratio.
Similar to the Golden Ratio,
the solution for the Silver Ratio also exhibits some interesting properties:
\begin{align}
\label{eq:8}
  x_{1}&= 1 +  \sqrt{2} = \sigma = 2.414 \\
  \sigma -2 &= \frac{1}{\sigma} = \frac{1}{1+\sqrt{2}}= \sqrt{2}-1 =0.414
\end{align}
Going back to the geometry of the unit circle with $r=1$, we see that the maximum length is 2.
Hence, the Silver Ratio can only be found after scaling down the reference distance $\overline{\mathrm{UM}}=1$ by a factor of 2.
This means the cardinal points move from $\mathrm{U}\to \mathrm{U}_{s}$ and $\mathrm{V}\to \mathrm{V}_{s}$.
\begin{figure}[htbp]
  \centering
  \includegraphics[width=0.3\textwidth]{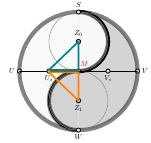}(a)\hfill
  \includegraphics[width=0.3\textwidth]{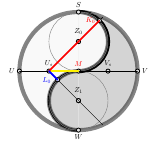}(b)\hfill
  \includegraphics[width=0.3\textwidth]{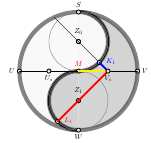}(c)
  \caption{\textsc{Yin}-\textsc{Yang} symbol (\textbf{YY}) with cardinal points (see \cref{tab:calendar}): (a) Right triangles to calculate distances, (b) Silver ratios $\sigma/2$ (red), $1/(2\sigma)$ (blue), \cref{eq:gs-4}, and reference segment $\overline{\mathrm{U_{s}M}}$ (yellow), (c) Same as (b), but for reference segment $\overline{\mathrm{V_{s}M}}$.}
  \label{fig:silvering}
\end{figure}

To illustrate the Silver Ratio in the \textbf{YY}, we use the auxiliary triangles $(\mathrm{U}_{s},\mathrm{M},\mathrm{Z}_{0})$  and $(\mathrm{U}_{s},\mathrm{M},\mathrm{Z}_{1})$ shown in \cref{fig:silvering}(a).
Because of $\overline{\mathrm{U}_{s}\mathrm{M}}=1/2$ and $\overline{\mathrm{Z}_{0}\mathrm{M}}=1/2$, $\overline{\mathrm{Z}_{1}\mathrm{M}}=1/2$, the Pythagorean theorem tells us that $\overline{\mathrm{U}_{s}\mathrm{Z}_{0}}=\overline{\mathrm{U}_{s}\mathrm{Z}_{1}}=\sqrt{2}/2$.
This gives us the distances of interest shown in \cref{fig:silvering}(b),  $\overline{\mathrm{U}_{s}\mathrm{K}_{0}}=1.207$ (in red) and $\overline{\mathrm{U}_{s}\mathrm{L}_{0}}=0.382$ (in blue), as follows: 
    \begin{equation}
    \label{eq:gs-4}
    \begin{aligned}
\overline{\mathrm{U_{s}K_{0}}}&=\overline{\mathrm{U_{s}Z_{0}}}+\overline{\mathrm{Z_{0}K_{0}}} = \frac{\sqrt{2}}{2}+\frac{1}{2}=\frac{\sigma}{2}  \\
      \overline{\mathrm{U_{s}L_{0}}}&=\overline{\mathrm{U_{s}Z_{1}}}-\overline{\mathrm{Z_{1}L_{0}}} = \frac{\sqrt{2}}{2}-\frac{1}{2}=\frac{\sigma-2}{2}=\frac{1}{2\sigma} 
    \end{aligned}
  \end{equation}
These two distances are related to the reference segment $\overline{\mathrm{U}_{s}\mathrm{M}}=1/2$ (in yellow).
Therefore the Silver Ratio appears both in the outer solution (in red) and in the inner solution (in blue).
The solutions for the reference point $\mathrm{V}_{s}$, shown in \cref{fig:silvering}(c), follow accordingly.

  \begin{figure}[htbp]
  \centering
  \includegraphics[width=0.5\textwidth]{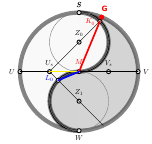}
  \caption{Interpretation of the Silver Ratio in the \textsc{Yin}-\textsc{Yang} symbol (\textbf{YY}): Definitions of $\mathrm{K}_{0}$, $\mathrm{L}_{0}$ and $\mathrm{G}$.}
  \label{fig:silver}
\end{figure}
  Similar to the procedure for the Golden Ratio, we now use the two points on the $\mathcal{S}$-curve to define the excess fractions (see \cref{fig:silver})
\begin{align}
  \label{eq:17}
  \rho_{K_{0}}&=\overline{\mathrm{M}\mathrm{K}_{0}} = 0.923 \nonumber \\
  \rho_{L_{0}}&=\overline{\mathrm{ML_{0}}} = 0.382 \nonumber \\
  \frac{\rho_{K_{0}}}{\rho_{L_{0}}}&=2.416 
\end{align}
Again, because of symmetries $\overline{\mathrm{ML_{0}}}=\overline{\mathrm{SK_{0}}}$,
the two ratios $\rho_{K_{0}}$ and $\rho_{L_{0}}$ form a right triangle in the inner circle, and
\begin{align}
  \label{eq:18}
  \rho_{K_{0}}^{2}+\rho_{L_{0}}^{2}=1
\end{align}
The last step to repeat is the calculation of the day number.
Starting from $\mathrm{M}$, we project the Silver Ratio point $\mathrm{K}_{0}$ onto the outer circle to obtain point $\mathrm{G}$ with an angle $\gamma=157.5^{\mathrm{o}}$ and a day number of 150.
This yields the Silver Ratio as follows:
\begin{align}
  \label{eq:19}
  \frac{a}{b}=\frac{365}{150}=2.433\;;\quad \frac{2a+b}{a}=\frac{880}{365}=2.410\;;\quad
  \frac{2a+b}{a}=\frac{520}{215}=2.418
\end{align}
All values are very close to $\sigma=2.414$.

Hence, we can confirm that the Silver Ratio has a physical interpretation analogous to that of the Golden Ratio,
both with respect to the duration of daytime and nighttime and to the division of the solar year into the number of days elapsed and the number of days remaining.

  \section{The Golden Spiral}
\label{sec:golden-spiral}

\begin{figure}[htbp]
  \centering
  \includegraphics[width=0.27\textwidth]{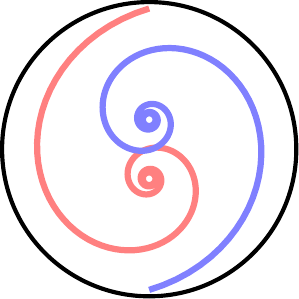}(a)\hfill
  \includegraphics[width=0.3\textwidth]{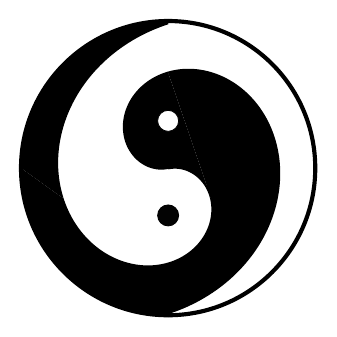}(b)\hfill
  \includegraphics[width=0.3\textwidth]{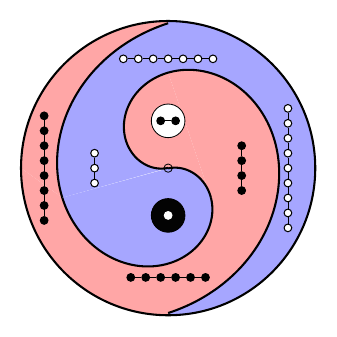}(c)
  \caption{(a) Golden spirals dividing the unit circle, (b) Area coloring to obtain the \textsc{Yin}-\textsc{Yang} symbol as an alternative to \cref{fig:tt}, (c) Combination of the \textsc{Yin}-\textsc{Yang} symbol with the \emph{Yellow River Map}, according to \citet{Berglund-1990-secretluoshu}.}
  \label{fig:spirals}
\end{figure}

The previous investigations are motivated by constructing the \textbf{YY} from nested circles, as shown in \cref{fig:facts}(a).
However, there are alternative construction methods that give a swirling appearance to the \textsc{Yin}-\textsc{Yang} symbol, as shown in \cref{fig:spirals}(b).
These are not based on \emph{circles}, but on \emph{spirals}.
Different types of spirals, e.g. Archimedean or Fibonacci spirals, give different patterns.
We illustrate the construction by using a \emph{Golden Spiral}
defined in polar coordinates as
\begin{align}
  \label{eq:21}
  r(\theta)=\phi^{\theta/(\pi/2)}
\end{align}
$r$ is the radius and $\theta$, measured in radians, is the angle defined counterclockwise. 
Here $\phi=1.618$ is the Golden Ratio already used before.
$\pi/2$ equals $90^{\mathrm{o}}$. 
That means that with every turn of $90^{\mathrm{o}}$, the radius increases by a factor of $\phi$.
We can rewrite \cref{eq:21} in the general form of a \emph{logarithmic} spiral:
\begin{align}
  \label{eq:20}
  r(\theta)= a\, e^{b\theta} 
\end{align}
where $a$ is the initial radius and $b$ is the growth factor. 
In \cref{fig:spirals}(a), 
we use $a=1$ and choose as starting points of two spirals the winter and the summer solstice, i.e. the bottom and the top of the unit circle. 
The growth factor $b$ in our case is chosen to be negative because our spirals shrink inward instead of growing outward.
To determine $b$, we combine \cref{eq:21} and \cref{eq:20} and choose $\theta=\pi/2$ as the reference point: 
\begin{align}
  \label{eq:22}
  \phi^{\theta/(\pi/2)}=e^{-b\theta} \;; \quad b= - \frac{\ln \phi}{\pi/2}=-0.306
\end{align}
It gives for the Golden Spirals shown in \cref{fig:spirals}(a), $r(\theta)=1.618^{-0.306\, \theta}$.
Now we color the areas between the outer circle and the spirals according to the \textbf{YY} in black and white, to obtain an alternative pattern for the \textbf{YY} as shown in \cref{fig:spirals}(b).

Compared to the construction based on circles, this pattern has several advantages.
First, while preserving the 50/50 distribution between black and white areas, this \textsc{Yin}-\textsc{Yang} symbol yields a more lively pattern, reminiscent of swirls or fish.
Second, the ``eyes'' appear naturally as the centers of the spirals.
Third, interestingly, the swirling pattern was used much earlier in history to symbolize the \textsc{Yin}-\textsc{Yang} principle than the circular pattern, which is quite modern.

Eventually, the spiral-based construction allows us to draw a nice connection to the so-called \textsc{He Tu}, the \emph{Yellow River Map}.
This magic figure appeared in the \emph{Book of Documents}
from about 500 years BCE \citep{Berglund-1990-secretluoshu}.
It presents in a square the numbers 1 to 10, symbolized by knotted strings with black and white knots (see \cref{fig:spirals}c).
Black refers to even numbers and to \textsc{Yin}, white to odd numbers and to \textsc{Yang}.
The distribution of these numbers is important.
Each of the four cardinal directions has two numbers assigned: \emph{South}, at the top,
has 7 and 2, \emph{North}, at the bottom, 6 and 1, \emph{East}, on the left side, 8 and 3, and \emph{West}, on the right side, has 9 and 4.
So each direction has one even and one odd number, and their difference is always 5.
The remaining numbers 5 and 10 are assigned to the center (\emph{not} shown in \cref{fig:spirals}c), where the knots for 5 form a cross in the middle and the knots for 10 surround it.

As \citet{Berglund-1990-secretluoshu} first reported, the two spirals separate the \emph{even} and \emph{odd} numbers in each direction such that all odd numbers are on one side and all even numbers on the other side.
This ultimately determines how the areas are colored:
The area with the even numbers (\textsc{Yin}) in black, the area with the odd numbers (\textsc{Yang}) in white.
To allow the numbers, i.e.\ the knotted strings, to be seen, I have colored them in red and blue instead.

To conclude, this construction yields an alternative pattern for the \textbf{YY}, which also captures a different symbolism.
Here the black and white areas do not refer to the excess of daytime or nighttime, but to number symbolism.
It is no small detail that the underlying \textsc{He Tu}, the Yellow River Map, is a \emph{square}, while the overlaying \textsc{Yin}-\textsc{Yang} symbol is a \emph{circle}.
The square represents earth, the circle represents heaven.
Both heaven and earth are themselves just different manifestations of \textsc{Yang} and \textsc{Yin}, forming an eternal unity.

\section{Discussion}
\label{sec:discussion}

\paragraph{Error estimates.}
\label{sec:numerical-errors}

It was \emph{not}
the aim of this paper to obtain the highest accuracy in comparison with empirical results. %
Nevertheless, because astronomical data exist, we can use them to estimate the errors of our calculations.
Before doing so, we summarize our simplifications in order to put the results in the right perspective.
In fact, the revolution of the Earth around the Sun and the rotation of the Earth are not as regular as we assumed.
The Earth's orbit is an ellipse, not a circle, the Earth itself is not a sphere but an ellipsoid, and its motion is influenced by the Moon and the gravitational forces of other planets.
Long-term dynamics such as the precession of the Earth's axis also play a role.
In short, it is a rather complex motion, which we have captured by the simplest possible rotational model based on only two parameters: the tilt of the Earth's axis, $\beta$, and the latitude of Zurich, $\lambda$.

Secondly, to simplify both the calculation and the visualization, we mapped the solar year, taken as the round value $365$, onto the circle of $360^{\mathrm{o}}$.
A quarter year was taken as a whole number of days, $91$, which introduces an error of 1 day; this discrepancy was assigned to the autumn equinox (172+91=263, but 355-91=264).
All calculated day numbers are rounded to whole days.
The dates of the cardinal points are uniformly set to the 21st day of the respective month, even though it is well known that they shift over long time periods (e.g.\ from the 21st of December to the 22nd of December).

With these simplifications in mind, let us take a look at the discrepancies in our calculations of daytime and nighttime duration.
According to \url{timeanddate.de}, the daytime duration for Zurich on 21 March 2026 was 12:12h, whereas we used 12:00h.
A value of 12:00h would have been reached between 17 and 18 March 2026.
The longest daytime for Zurich occurs on 21 June, with a length of 15:57h, whereas we calculated 15:45h.
Likewise, the shortest daytime on 21 December is 8:27h, while we obtained 8:15h.
For 24 April we estimated the daytime as 13:52h, whereas the tabulated value is 14:02h,
and for 24 November we obtained 8:42h, against a tabulated value of 8:52h.
Deviations are thus present as expected, but the relative error is around 0.01, or 1\%.

\paragraph{Relations to the \textbf{YY} .}
We have demonstrated that (i) the \textbf{YY}  has a \emph{physical} interpretation in terms of the daytime/nighttime dynamics and (ii) the correct appearance of the smaller circles and their coloring can be derived from a simple model whose only free parameter is the latitude of Zurich.

Because \textsc{Yin} and \textsc{Yang} have always been associated with the sunlit and shadowed areas of, e.g., a valley, the connection to the daylight problem may seem obvious.
Nevertheless, \emph{deriving} such a relation is an entirely different matter from offering a metaphysical interpretation.
Key to our derivations was the \emph{reduced variable} $\rho(\gamma)$, called the excess fraction of daytime/nighttime, which yields an analytic expression for the $\mathcal{S}$-curve dividing the black and white regions.

\paragraph{Relations to the Golden Ratio.}

The physical interpretation of the Golden and Silver ratios appearing in the geometry of the \textbf{YY}  was an unexpected bonus.
We were able to demonstrate that these ratios divide the solar year correctly into days elapsed and days remaining, and can thus be given a calendrical interpretation.
At second glance, the correct proportions may seem less surprising, since one can always arbitrarily divide a length $L$ so as to obtain these ratios.
However, as we have verified, the \emph{angles} $\gamma$ defined with respect to the winter solstice would \emph{not} yield such relations.
These relations emerge for the day numbers only after correcting the angles by the appropriate offset of $\gamma=11^{\mathrm{o}}$ or $t=10$ days.
What is remarkable, therefore, is that this offset coincides precisely with the start of the New Year according to the Gregorian calendar, something that was by no means expected.

\paragraph{Relations to the ``Book of Changes''.}
The \textsc{Yijing}, the more than 2500-year-old classic \emph{Book of Changes}, already employs the concept of \textsc{Yin} and \textsc{Yang} to describe the fundamental structure of reality as a \enquote{Supreme Polarity}.
\textsc{Yin} is associated with \emph{even} numbers, in particular 6 and 8, and \textsc{Yang} with \emph{odd} numbers, in particular 7 and 9 \citep{Wilhelm-Baynes-1977-iching}.

The \textsc{Yijing} adds a \emph{dynamics} to this structure.
\textsc{Yin} and \textsc{Yang} each appear in two states, one immutable and one changeable.
Specifically, \enquote{old} \textsc{Yin}, characterized by the number 6, is able to transform into \enquote{young} \textsc{Yang}, characterized by the number 7.
Likewise, \enquote{old} \textsc{Yang}, characterized by the number 9, is able to transform into \enquote{young} \textsc{Yin}, characterized by the number 8.

Our construction of the \textbf{YY} with its four quadrants (\cref{fig:filled}) connects these numbers to their respective quadrants as follows: Quadrant~I corresponds to \emph{young} \textsc{Yin}, 8, because it has not yet reached the critical threshold for change.
Quadrant~II corresponds to \emph{old} \textsc{Yin}, 6, because we observe its transformation into \textsc{Yang}.
Quadrant~III corresponds to \emph{young} \textsc{Yang}, 7, which has just appeared and has not yet grown to a critical level.
Quadrant~IV corresponds to \emph{old} \textsc{Yang}, 9, because it is in the process of being transformed into \emph{young} \textsc{Yin}.

We particularly highlight that this process of transformation follows a sequence of \emph{activity} (\enquote{change}) and \emph{tranquility} (\enquote{rest}).
The latter does not imply that nothing happens.
In fact, a continuous process unfolds \enquote{under the hood}, but the system \enquote{holds still} because its time has not yet come.
Thus, the dynamics we observe in our daytime model perfectly reflect the intentions of the \textsc{Yijing}.

\paragraph{Relations to phase transitions.}

It is possible to relate the \textsc{Yin}-\textsc{Yang} dynamics to phenomena such as phase transitions in physics, bifurcations in nonlinear dynamics, or regime shifts in ecology.
The \emph{control parameter} is the solar declination $\delta(t)$, the angle between the ecliptic and the equator.
It varies over time from $\delta(t)=-23.44^{\mathrm{o}}$ at the winter solstice to $\delta(t)=23.44^{\mathrm{o}}$ at the summer solstice (see also \cref{tab:calendar}).
This change is driven by an external process, the Earth's revolution around the Sun, which gives rise to the four seasons.
When the control parameter passes through a critical value, $\delta(t)=0$, a transition is induced because the stability of the current state is lost.
This critical value $\delta(t)=0$ occurs at the vernal and autumnal equinoxes.
Consequently, the transition takes place in the II. and IV. quarters of the year.
In these two quarters we observe the $\mathcal{S}$-curve, i.e.\ the coexistence of black and white.

The new stable state is reached at the summer or winter solstice, respectively.
Thus, during the I. and III. quarters the dynamical system is stable.
More precisely, it is \emph{metastable}, since the ratio between \textsc{Yin} and \textsc{Yang} is continuously changing.
However, during the I. and III. quarters this change has not yet reached a critical level.
Therefore, the I. quarter is colored entirely \emph{black} and the III. entirely \emph{white} in the \textbf{YY}, indicating stability.

\vspace{3ex}
To conclude, reconstructing the \textsc{Yin}-\textsc{Yang} symbol from a model of daytime change may not add entirely new insights into the \emph{symbol} itself.
This was not to be expected, given the extensive literature on this topic spanning hundreds of years.
It is nevertheless surprising to observe natural processes at work even in such a famous and iconic pattern.
More importantly, the findings presented here are perfectly aligned with the established concepts of Chinese philosophy, shedding new light on the intricate connection between natural processes and systems of thought.

\end{document}